\documentclass[11pt]{article}
\usepackage{epsfig}

\newcommand{\beq}{\begin{equation}}
\newcommand{\eeq}{\end{equation}}
\newcommand{\beqa}{\begin{eqnarray}}
\newcommand{\eeqa}{\end{eqnarray}}

\newcommand{\gaem}{\stackrel{>}{\sim}}

\newcommand{\tr}{{\rm Tr}}

\begin{document}

\begin{titlepage}
\def\thepage {}        

\title{Large-$N$ and Vacuum Alignment in Topcolor Models}

\author{
R. Sekhar Chivukula\thanks{sekhar@bu.edu}\\
Boston University \\
Boston MA  02215
\and
Howard Georgi\thanks{georgi@physics.harvard.edu}\\
Harvard University\\
Cambridge, MA 02138}

\date{May 27, 1998}

\maketitle

\bigskip
\begin{picture}(0,0)(0,0)
\put(295,250){BUHEP-98-12}
\put(295,235){HUTP-98/A025}
\end{picture}
\vspace{24pt}

\begin{abstract}

  Topcolor and topcolor-assisted technicolor provide examples of
  dynamical electroweak symmetry breaking which include
  top-condensation, thereby naturally incorporating a heavy top quark.
  In this note we discuss the roles of the Nambu-Jona-Lasinio (NJL) and
  large-$N$ approximations often used in phenomenological analyses of
  these models. We show that, in order to provide for top-condensation
  but not bottom-condensation, the top-color coupling must be adjusted
  to equal the critical value for chiral symmetry breaking up to ${\cal
    O}(1/N)$ in any theory in which the isospin-violating ``tilting''
  interaction is a $U(1)$ gauge interaction. A consequence of these
  considerations is that the potentially dangerous ``bottom-pions'' are
  naturally light. We also show that the contributions to $\rho-1$
  previously estimated are of leading-order in $N$, are not included in
  the usual NJL analysis, and are the result of ``vacuum-alignment''.

\pagestyle{empty}
\end{abstract}
\end{titlepage}

\setcounter{section}{0}

\section{Introduction}
\setcounter{equation}{0}

Topcolor and topcolor-assisted technicolor
\cite{Hill:1991at,Hill:1995hp} provide examples of dynamical electroweak
symmetry breaking which include top-condensation
\cite{Miranskii:1989ds,Miranskii:1989xi,Nambu:1989jt,Marciano:1989xd,Bardeen:1990ds,Cvetic:1997eb},
thereby naturally incorporating a heavy top quark. In this note we
discuss the roles of the Nambu--Jona-Lasinio (NJL) and large-$N$
approximations often used in phenomenological analyses of these models.

We begin by reviewing the usual analysis of topcolor in the large-$N$
NJL approximation.  From this analysis one finds immediately
\cite{Kominis:1995fj} that, in order to provide for top-condensation but
not bottom-condensation, the top-color coupling must be adjusted to
equal the critical value for chiral symmetry breaking up to ${\cal
  O}(1/N)$. We show that this is a general result in any theory in which
the isospin-violating ``tilting'' interaction is a $U(1)$ gauge
interaction. A consequence of these considerations is that the
potentially dangerous ``bottom-pions''
\cite{Kominis:1995fj,Balaji:1996rh} are naturally light. In the NJL
approximation, the effects of the portion of the strong topcolor
interactions coupling left-handed currents to right-handed currents are
presumed to dominate, and to give rise to chiral symmetry breaking. We
show that other portions of the topcolor interactions, in particular
products of pairs of left-handed currents or right-handed currents, give
rise to contributions to $\rho-1$. These contributions, which have been
estimated previously \cite{Chivukula:1995dc} by direct computation, are
of leading-order in $N$ and are the result of ``vacuum-alignment''
\cite{Dashen:1971et,Weinberg:1976gm,Peskin:1980gc,Preskill:1980mz}.

\section{Topcolor}
\setcounter{equation}{0}

In top-color models \cite{Hill:1991at,Hill:1995hp} all or part of
electroweak symmetry breaking is due to the presence of a
top-condensate.  In many models this condensate is driven by the
combination of a strong isospin-symmetric top-color interaction and an
additional isospin-breaking $U(1)$ gauge boson which couples only to the
third generation of quarks. These additional interactions are strong,
but are spontaneously broken at a scale $M \gaem 1$ TeV.  In the
simplest model \cite{Hill:1995hp}, the couplings of the third generation
of quarks to the new $U(1)$ interaction were taken to be proportional to
weak hypercharge, and the masses of the heavy topcolor and $U(1)$ gauge
bosons were taken to be comparable. At low energies, the top-color and
hypercharge interactions of the third generation of quarks could then be
approximated by the four-fermion operators
\beq
{\cal L}_{4f} = -{{4\pi
\kappa_{tc}}\over{M^2}}\left[\overline{\psi}\gamma_\mu {{\lambda^a}\over{2}}
\psi \right]^2
-{{4\pi \kappa_1}\over{M^2}}\left[{1\over3}\overline{\psi_L}\gamma_\mu  \psi_L
+{4\over3}\overline{t_R}\gamma_\mu  t_R
-{2\over3}\overline{b_R}\gamma_\mu  b_R
\right]^2~,
\label{L4t}
\eeq
where $\psi$ represents the top-bottom doublet, $\kappa_{tc}$ and
$\kappa_1$ are related respectively to the top-color and $U(1)$
gauge-couplings squared.

The analysis of the dynamics of this model usually proceeds in two
steps. First, using a Fierz transformation, those (LR) terms in 
eqn.~(\ref{L4t}) above which couple left-handed and right-handed currents and
can be converted into products of color-singlet
scalar/pseudoscalar bilinears are rewritten as
\beq
{\cal L}_{NJL} = +{8\pi\over{M^2}}
\left[\kappa_t (\overline{\psi}_L t_R)(\overline{t}_R \psi_L)
+\kappa_b (\overline{\psi}_L b_R)(\overline{b}_R \psi_L)\right]
~,
\label{LNJL}
\eeq
where
\beq
\kappa_t = \kappa_{tc} + {8\kappa_1 \over 9N} \ \ \ \ \ \ 
\kappa_b = \kappa_{tc} - {4\kappa_1 \over 9N}~,
\label{kappas}
\eeq
and $N$(=3) is the number of top-colors.  

Next, this effective ``NJL'' model (eqn. (\ref{LNJL}))
\cite{Nambu:1961er,Miranskii:1989ds,Miranskii:1989xi,Nambu:1989jt,Marciano:1989xd,Bardeen:1990ds}
is analyzed to leading order in $N$. This can be conveniently done by
introducing a complex $2 \times 2$ matrix field $\Phi$ and writing the
NJL interactions in the form
\beq
{\cal L}_{NJL} \to \left(\overline{\psi}_L \Phi \left(\matrix{t_R \cr
b_R \cr}\right) + h.c.\right) - {M^2 \over 8\pi} \tr \left[ \Phi^\dagger \Phi
\left(\matrix{{1\over \kappa_t} & 0\cr 0& {1\over \kappa_b}\cr}\right)\right]~.
\label{Lscalar}
\eeq
Note that, written this way, the interaction of $\Phi$ with the fermions
is $SU(2)_L \times SU(2)_R$ symmetric.  To leading order in $N$, the
theory is now solved by computing the trace of the fermion propagator in
the presence of a background $\Phi$ field \cite{Bardeen:1990ds}.
Computing the effective potential for the field $\Phi$ using a
momentum-space cutoff of order $M$, we find
\beq
V^{eff}(\Phi) \approx  {M^2\over 8\pi\kappa_c} \tr \left[\Phi^\dagger \Phi
\left(\matrix{{\kappa_c \over \kappa_t}-1 & 0\cr 0& 
{\kappa_c \over \kappa_b}-1\cr}\right)\right]
+ {N \over 16\pi^2}\tr\left[(\Phi^\dagger\Phi)^2
\log\left({M^2 \over \Phi^\dagger\Phi}\right)\right]
~,
\label{veff}
\eeq
where $\kappa_c = \pi/N$.

The essential features of this model in the large-$N$ NJL approximation
can now be determined from eqn. (\ref{veff}). For $\kappa_{t,b}$ close to
$\kappa_c$, the field $\Phi$ yields four light complex scalar fields
which have the quantum numbers of two independent 2-component ``Higgs''
fields $(\phi_t\ \&\ \phi_b)$ (with hypercharges $\mp 1$, respectively).
Choosing the values of $\kappa_{tc}$ and $\kappa_1$
such that
\beq
\kappa_t\, >\, \kappa_c\, >\, \kappa_b~,
\eeq
we obtain the phenomenologically desirable result that the
doublet $\phi_t$ develops a vacuum expectation value, 
\beq
\langle \phi_t \rangle = \left(\matrix{{f_t\over \sqrt{2}} \cr 0
\cr}\right)~,
\eeq
giving rise to a (potentially large) top-quark mass and leaving $\langle
\phi_b \rangle \equiv 0$.  Taking into account the necessary
wavefunction renormalization for the scalar field \cite{Bardeen:1990ds}
we find \cite{Pagels:1979hd}
\beq
f^2_t \approx {N\over 8\pi^2}m^2_t\log\left({M^2\over m^2_t}\right)~.
\eeq
For $m_t \approx 175$ GeV and $M \approx 1$ TeV, this yields $f_t
\approx 64$ GeV.  From the equations of motion for $\phi_t$ derived from
eqn. (\ref{Lscalar}), we see that this expectation value can be
interpreted as a top-quark condensate. To the extent that $\kappa_t$ and
$\kappa_b$ are close to the ``critical value'' $\kappa_c$ for chiral
symmetry breaking, the fields $\phi_t$ and $\phi_b$ have masses (and
expectation values) small compared to $M$ and are composite Higgs fields
\cite{Bardeen:1990ds,Hasenfratz:1991it}.

At this level of approximation, for fixed $M$, the effective
mass-squareds of the Higgs fields change smoothly from positive to
negative as the respective $\kappa$'s vary from below to above the
``critical value'' $\kappa_c$.  That is, in this approximation the
chiral phase transition (for fixed $M$ and viewed as a function of
$\kappa_{tc}$) is second-order. For $\kappa$'s close to $\kappa_c$ the
effective scalar lagrangian is a Landau-Ginzburg theory of the chiral
phase transition with order parameter $\Phi$.  

\section{Large-$N$ and the Chiral Phase Transition}
\setcounter{equation}{0}

From eqn. (\ref{kappas}), we see that the difference $\kappa_t-\kappa_b$
is of order $\kappa_1/N$. We will argue shortly that the ratio
$\kappa_1/\kappa_{tc}$ is
independent of $N$. Therefore, in order to provide for
top-condensation but not bottom-condensation, the top-color coupling
must be adjusted to equal the critical value for chiral symmetry
breaking up to ${\cal O}(1/N)$. In this section we show that this is a
general property of the large-$N$ limit, independent of the NJL
approximation. We will also see that it is independent of the assumption that
the topcolor and strong $U(1)$ gauge boson masses are equal --- it persists
even if these masses are very different.

\begin{figure}
\begin{center}
\epsfig{file=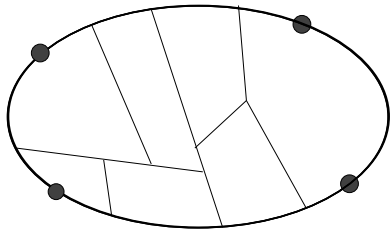,height=3cm}
\end{center}
\caption{A ``typical'' planar contribution, of
leading order in $1/N$, to the effective
potential for $\Phi$. Fermions are represented by the thick
solid lines, gluons by the thin solid lines, and insertions
of $\Phi$ by the dots.}
\label{figone}
\end{figure}

Consider the topcolor theory in large-$N$
\cite{'tHooft:1974jz}.  As in QCD, in order to have a well-defined
high-energy theory, we must choose a topcolor coupling
\beq
g_{tc} = {\tilde{g}_{tc}\over \sqrt{N}}~,
\label{gtcN}
\eeq
and hold $\tilde{g}_{tc}$ fixed as $N \to \infty$. 
The chiral symmetries of the topcolor theory are $SU(2)_L \times
SU(2)_R$, under which the left-handed top-bottom doublet transforms as a
$(2,1)$ and the right-handed top and bottom transform together as a
$(1,2)$.  The behavior of the chiral symmetries is governed by the
effective potential for an order parameter $\Phi$, which transforms as a
$(2,\tilde{2})$ under $SU(2)_L \times SU(2)_R$. To leading order in
$1/N$, this potential comes from the sum of all {\it planar} diagrams
involving one fermion loop as shown in figure \ref{figone}, and is
${\cal O}(N)$.  The flavor structure of this class of diagrams insures
that the vectorial subgroup, $SU(2)_V$, remains
unbroken\footnote{In fact, for any QCD-like vector gauge theory, this
remains true exactly \cite{Vafa:1984tf}.} \cite{Coleman:1980mx}.

Because $g_{tc}$ varies with $N$, we must be careful about what we
mean by the scale of topcolor breaking. It will be most useful to define
this scale as the Lagrangian mass of the topcolor gauge boson. This is
appropriate because it is this mass which acts as the cut-off for the
effective theory below the symmetry breaking scale. In fact we will soon
see that the topcolor boson mass, $M$, remains fixed as
$N\rightarrow\infty$, which means the the vacuum expectation value that
is responsible for the breaking must actually grow like $\sqrt N$.

Let $\Lambda_{tc}$ be the scale at which the topcolor interactions would
become strong if topcolor remained unbroken, {\it i.e.} it is the analog
of $\Lambda_{QCD}$ for the ordinary strong interactions. $\Lambda_{tc}$
is then independent \cite{'tHooft:1974jz} of $N$ as $N\rightarrow\infty$. If
topcolor is broken at scale $M$, we can analyze the theory in two
limits~\cite{Chivukula:1990bc}. First, if $M \gg \Lambda_{tc}$, we
expect the low-energy theory to contain massless fermions which interact
(ignoring the standard model interactions) only by the exchange of
heavy, weakly-coupled, topgluons. In this limit, chiral symmetry is
unbroken and $\langle \Phi \rangle = 0$. On the other hand, if $M \ll
\Lambda_{tc}$ the topcolor interactions become strong and we may expect
chiral symmetry to be broken in a manner similar to that in QCD. Here
chiral symmetry is broken and $\langle \Phi \rangle \propto I \neq 0$.
If the transition between these two regimes is continuous, {\it i.e.} if
the chiral-symmetry breaking transition is of second order as
$g_{tc}^2(M)$ (the value of the topcolor coupling at scale $M$) is
varied, then the mass-squared of the effective $\Phi$ field goes
continuously through zero at a critical coupling $g_c^2$.  

Since $\Lambda_{tc}$ is independent of $N$ as $N \rightarrow \infty$, $M$ must
also be independent of $N$ as $N \rightarrow\infty$.  For this to be consistent
with the condition of criticality, $g_{tc}^2(M)=g_c^2$, $g_c^2$ must be
proportional to $1/N$, {\it i.e.}
\beq
g^2_c = {{\tilde g}^2_c \over N}~,
\eeq
with ${\tilde g}_c$ of ${\cal O}(1)$.  This agrees with the NJL analysis
in which $\kappa_c=\pi/N$ in (\ref{veff}) is proportional to $g_c^2$.

If the transition is second-order, the effective low-energy theory for
$g_{tc}^2(M)$ close to $g_c^2$ is one with a light scalar $\Phi$
(with mass $\ll M$) coupled to fermions, just as in the NJL analysis
 \cite{Chivukula:1990bc}.
Therefore, while the analysis of topcolor presented in the previous
section depends on both the large-$N$ and NJL approximations to the full
top-color theory, the important properties may be expected to survive
beyond these approximations so long as the chiral phase transition is
sufficiently second-order \cite{Chivukula:1990bc}.

\begin{figure}
\begin{center}
\epsfig{file=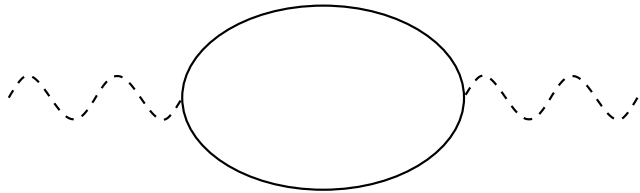,height=2.5cm}
\end{center}
\caption{Fermion-loop contribution to vacuum polarization
correction to the $U(1)$ gauge-boson propagator. The corresponding
term in the $\beta$-function for the coupling
grows as $N$. To have a well-defined large-$N$ limit, therefore, 
the $U(1)$ coupling must scale as $N^{-1/2}$.}
\label{figtwo}
\end{figure}

\begin{figure}
\begin{center}
\epsfig{file=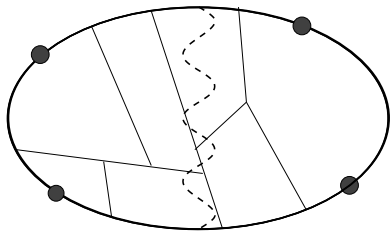,height=3cm}
\end{center}
\caption{A typical contribution to the effective potential
  for $\Phi$ to leading order in $1/N$ and lowest order in the
  (symmetry-violating) $U(1)$ couplings. The $U(1)$ gauge boson
  propagator is shown in dashed lines.}
\label{figthree}
\end{figure}

Now consider the contribution of the isospin-violating $U(1)$
interaction. In order to have a well-defined large-$N$ limit, we must
insure that the vacuum polarization diagrams (see figure \ref{figtwo})
have a finite large-$N$ limit. Hence, in analogy with eqn. (\ref{gtcN}),
we must take a $U(1)$ coupling
\beq
g_1 = {{\tilde g}_1\over \sqrt{N}}~,
\eeq
and hold $\tilde{g}_1$ fixed as $N \to \infty$. This immediately implies that
$\kappa_1$ in (\ref{L4t}), is proportional to $1/N$ ---
so that $\kappa_1/\kappa_{tc}$ in independent of $N$, as promised above.

We can now generalize the analysis to the general effective field theory
description, beyond the NJL approximation. To leading order in
$1/N$, the contribution of the $U(1)$ interaction to the effective
potential for $\Phi$ comes from planar diagrams involving one fermion
loop and one $U(1)$ gauge-boson exchange (figure \ref{figthree}).
Regardless of the specific charges chosen, these contributions are ${\cal
  O}(1)$.

For the $U(1)$ interaction to ``tilt'' the vacuum and break $SU(2)_V$,
the contribution of the $U(1)$ coupling to the mass-squared of the
$\Phi$ field must compete with the leading contribution from topcolor.
Therefore, the contribution of the topcolor interactions to the
mass-squared of $\Phi$ must be adjusted to be ${\cal O}(1)$. That is,
the topcolor chiral phase transition must be a second order transition
to subleading order\footnote{It is of note that at next-to-leading order in
  $N$ there are contributions which may make the phase transition weakly
  first order \cite{Chivukula:1993pm,Bardeen:1994pj} due to the
  Coleman-Weinberg mechanism \cite{Coleman:1973tz}.} in $N$ {\it and}
\beq
{\Delta g^2_{tc}(M) \over g^2_c} = 
{g^2_{tc}(M)-g^2_c \over g^2_c} = {\cal O}\left({1\over N}\right)\, ,
\eeq
{\it i.e.} ${\tilde g}^2_{tc}(M)$ must be ``tuned'' to equal ${\tilde
  g}^2_c$ to ${\cal O}(1/N)$.

These considerations have an immediate phenomenological consequence.
Treating $\Phi = (\phi_t\, \phi_b)$ as a pair of Higgs fields, we see
that the difference in the mass-squared of $\phi_t$ and $\phi_b$
\beq
{|m^2_{\phi_t} - m^2_{\phi_b}|\over M^2} = {\cal O}\left( {1 \over
N}\right)
\eeq
is subleading\footnote{This can be seen in eqn. \ref{veff}, for example,
because $m^2_{\phi_t} - m^2_{\phi_b} \propto |\kappa_t - \kappa_b| =
 {\cal O}(1/N)$.} in $N$ \cite{Kominis:1995fj,Buchalla:1996dp}.
Independent of the NJL-approximation, therefore, we expect that the mass
of the $\phi_b$ doublet cannot be significantly larger than the weak
scale unless $M$ is taken to be much larger than the weak scale. This
extra light $\phi_b$ doublet could give rise to dangerous effects in
$B$-$\bar{B}$ mixing and $B$-meson decays
\cite{Kominis:1995fj,Balaji:1996rh}.

Note that this argument does not depend on any of the details that lead to
(\ref{L4t}). We need not assume that the topcolor and $U(1)$ gauge boson
masses are equal, or even of the same order of magnitude.

\section{Beyond the NJL Approximation: $\Delta \rho$}
\setcounter{equation}{0}

The NJL approximation treats only the terms in eqn. (\ref{L4t}) which
couple left-handed and right-handed currents. Consider the effects of
the topcolor interactions in eqn. (\ref{L4t}) coupling {\it pairs} of
left-handed (LL) or right-handed (RR) currents, which are not included
in the NJL approximation.  The LL terms may be Fierz transformed to
color-singlet form
\beq
\left(\bar{\psi}_L\gamma^\mu {\lambda^a\over 2} \psi_L\right)^2
\to {1\over 2} (\bar{\psi}_L \gamma^\mu \psi_L)^x_y
(\bar{\psi}_L \gamma_\mu \psi_L)^y_x - 
{1\over 2N} (\bar{\psi}_L \gamma^\mu \psi_L)
(\bar{\psi}_L \gamma_\mu \psi_L)~.
\label{llint}
\eeq
Here $x$ and $y$ are $SU(2)_L$ (flavor) indices and an analogous
expression holds for the terms involving a product of right-handed
currents with $L \leftrightarrow R$.

\begin{figure}
\begin{center}
\epsfig{file=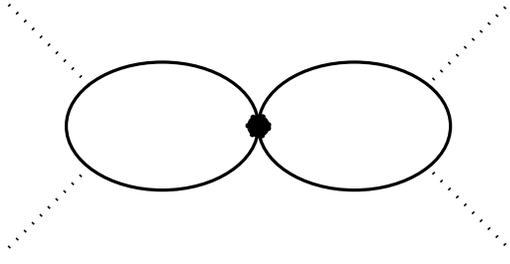,height=3.5cm}
\end{center}
\caption{Leading contribution of product of left-handed
currents in eqn. (\ref{L4t}) to the low-energy effective
scalar theory. Fermion lines are solid, scalar ($\Phi$) lines
are dotted, and the left-handed four-fermion operator is represented
by the solid circle in the center.}
\label{figfour}
\end{figure}

Treating the operator in eqn. (\ref{llint}) as a perturbation, its
contribution in the low-energy effective scalar theory may be computed
to leading-order as shown in figure \ref{figfour}. As in
QCD \cite{Bardeen:1986vp,Chivukula:1986du}, to leading-order
in $N$, the ``vacuum insertion'' approximation may be used to
evaluate this contribution. The left-handed and
right-handed flavor currents
can then be matched individually to the corresponding flavor
currents in the low-energy scalar theory. Since $\Phi$ transforms
as a $(2,\bar{2})$ under $SU(2)_L \times SU(2)_R$, we find
{\renewcommand{\arraystretch}{2}
\beq
\begin{array}{l}
\displaystyle
(\bar{\psi}_{L} \gamma^\mu \psi_{L})^x_y \to - i(\Phi 
\stackrel{\leftrightarrow}{\partial^\mu} \Phi^\dagger)^x_y
+ {\cal O}\left({\Phi\Phi^\dagger \over M^2},{\partial^2\over M^2}\right)~,\\
\displaystyle
(\bar{\psi}_{R} \gamma^\mu \psi_{R})^x_y \to - i(\Phi^\dagger 
\stackrel{\leftrightarrow}{\partial^\mu} \Phi)^x_y
+ {\cal O}\left({\Phi\Phi^\dagger \over M^2},{\partial^2\over M^2}\right)~.
\end{array}
\label{matching}
\eeq
}
The fermion currents above contains both $SU(2)_{L,R}$ {\it and}
$U(1)_{L,R}$ pieces.  However, in the {\it linear} sigma model, the
lowest-order term in the effective Lagrangian which distinguishes
between the $SU(2)$ triplet and singlet pieces of $\Phi$ is of dimension
six. Therefore the relation of eqn. (\ref{matching}) holds for both parts
up to corrections of ${\cal O}(1/M^2)$.

In the leading vacuum-insertion approximation we may then
immediately write down the effective interaction arising from
the left-handed and right-handed operators of the form of
eqn. (\ref{llint}):
{\renewcommand{\arraystretch}{2}
\beq
\begin{array}{l}
\displaystyle
\delta {\cal L}_{LL} = +\,{2\pi\kappa_{tc} \over M^2}
\tr(\Phi\stackrel{\leftrightarrow}{\partial^\mu} \Phi^\dagger)
(\Phi\stackrel{\leftrightarrow}{\partial_\mu} \Phi^\dagger)
~,\\
\displaystyle
\delta {\cal L}_{RR} = +\,{2\pi\kappa_{tc} \over M^2}
\tr(\Phi^\dagger\stackrel{\leftrightarrow}{\partial^\mu} \Phi)
(\Phi^\dagger\stackrel{\leftrightarrow}{\partial_\mu} \Phi)
~,
\end{array}
\label{effdrho}
\eeq
}
plus corrections of order $M^{-4}$ and $1/N$. Note that these interactions
are $SU(2)_L \times SU(2)_R$ invariant. This is as it should
be since the topcolor interactions are themselves $SU(2)_L\times
SU(2)_R$ invariant.

The vacuum expectation value of the field is
\beq
\langle \Phi \rangle = 
\left(
\begin{array}{cc}
{f_t\over \sqrt{2}} & 0 \cr
0 & 0 
\end{array}\right)~,
\label{vev}
\eeq
which breaks $SU(2)_L \times SU(2)_R \to U(1)_{em}$. Note that the
potential ``custodial'' $SU(2)_V$ \cite{Sikivie:1980hm} symmetry is
broken by the alignment of the vacuum
\cite{Dashen:1971et,Weinberg:1976gm,Peskin:1980gc,Preskill:1980mz}.
Because of the usual {\it accidental} symmetry, the terms of dimension
four or less in the effective Lagrangian {\it cannot} give rise to a
deviation of the rho parameter,
\beq
\rho = {M^2_W \over M^2_Z \cos^2\theta_W},
\eeq
from one. The leading contribution arises from the operator
in eqn. (\ref{effdrho}). Promoting the partial derivatives to
gauge covariant derivatives, we may calculate the gauge-boson
masses and find the contribution to $\rho - 1$ 
\beq
\Delta \rho = {2\pi e^2\kappa_{tc} \over \sin^2\theta_W \cos^2\theta_W}
{f^4_t \over M^2_Z M^2}~.
\label{drho}
\eeq
Note that this contribution to $\Delta\rho$ is of leading-order in $N$
and results from the LL and RR topcolor interactions not included in the
NJL approximation.  To leading order in $N$, this result agrees with the
calculation given in \cite{Chivukula:1995dc}.

The calculation of $\Delta\rho$ relies on treating the LL and RR
portions of the topcolor interactions as perturbations to the NJL model.
However, even beyond this approximation we expect the effective
Lagrangian for the composite $\Phi$ will contain terms of the form shown
in eqn. (\ref{effdrho}). We therefore generally expect contributions of
this order of magnitude\footnote{This may be viewed as a special case of
  the constraints on composite Higgs models discussed in
  \cite{Chivukula:1996sn}.}.

\begin{figure}
\begin{center}
\epsfig{file=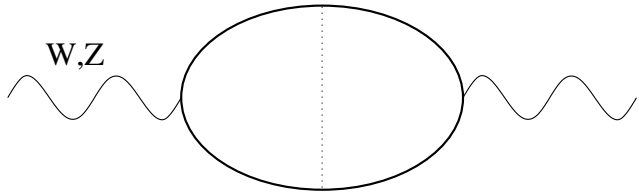,height=2.5cm}
\end{center}
\caption{Example of a potentially large two-loop contribution
to $\Delta\rho$ arising from exchange of light composite
scalars. Fermions (the top and bottom) are represented by
solid lines and scalars by the dotted line.
These contributions are subleading in $1/N$,
but are not suppressed by $1/M^2$.}
\label{figfive}
\end{figure}

The corrections discussed above are of leading-order in $1/N$, but are
suppressed by $1/M^2$. There are also corrections coming from the
exchange of the light scalars in $\Phi$ (such as that shown in fig.
\ref{figfive}). These contributions are subleading in $1/N$ but are {\it
  not} suppressed by $1/M^2$. They are also potentially large because
the couplings to the top-quark are of order $m_t/f_t$, approximately
four times {\it larger} than the corresponding contributions from a
standard model Higgs boson. It is amusing to note that if these
particles were light, this contribution could be {\it negative}
\cite{vanderBij:1987hy,Barbieri:1992nz,Barbieri:1992dq,Fleischer:1995cb}.
However, phenomenological constraints from $Z \to b\bar{b}$
\cite{Burdman:1997pf} require that these particles be relatively heavy
and these contributions are probably suppressed.

Finally, we comment on the situation in models
\cite{Appelquist:1996xx,Buchalla:1996dp} which do {\it not} contain a
$U(1)$ tilting interaction. Instead, the strong topcolor group is
arranged to couple to the left-handed top-bottom doublet (and, in
``topcolor II'' \cite{Buchalla:1996dp}, the charm-strange doublet) and
the right-handed top, but not to the right-handed bottom. The large-$N$
analysis of the first section then no longer applies --- however,
anomaly cancellation implies that the number of {\it flavors} to which
topcolor couples must scale with the number of colors. This situation is
reminiscent of the large-$N$ \& large-$N_f$ analysis of
\cite{Veneziano:1976wm}. Such a theory will have a large number of
scalars and contributions of the form shown in fig.  \ref{figfive} would
no longer be suppressed by $1/N$.  Furthermore, as the topcolor
interactions themselves violate custodial $SU(2)$, there will still
generically be contributions analogous to eqn.  (\ref{effdrho}) which
yield contributions to $\Delta\rho$ of the same order of magnitude
\cite{Chivukula:1996sn}.

\bigskip

Topcolor and topcolor-assisted technicolor
\cite{Hill:1991at,Hill:1995hp} provide examples of dynamical electroweak
symmetry breaking which include top-condensation
\cite{Miranskii:1989ds,Miranskii:1989xi,Nambu:1989jt,Marciano:1989xd,Bardeen:1990ds,Cvetic:1997eb},
thereby naturally incorporating a heavy top quark. In this note we have
discussed the roles of the Nambu--Jona-Lasinio and large-$N$
approximations used in phenomenological analyses and have discussed the
dynamical behavior of topcolor theories beyond these approximations. We
have shown that, in models with a $U(1)$ tilting interaction, in order
to provide for top-condensation but not bottom-condensation the
top-color coupling must be adjusted to equal the critical value for
chiral symmetry breaking up to ${\cal O}(1/N)$.  A consequence of these
considerations is that the potentially dangerous ``bottom-pions''
\cite{Kominis:1995fj,Balaji:1996rh} are naturally light. We have also
considered contributions beyond the NJL approximation and shown that the
other portions of the topcolor interactions, in particular products of
pairs of left-handed currents or right-handed currents, give rise to
contributions to $\rho-1$. These contributions, which have been
estimated previously \cite{Chivukula:1995dc} by direct computation, are
of leading-order in $N$ and are the result of vacuum-alignment.

\bigskip


\centerline{\bf Acknowledgments}

We thank Bogdan Dobrescu, Nick Evans, Chris Hill, Elizabeth Simmons, and
John Terning for discussions and comments on the manuscript. Portions of
this work were completed while R.S.C. visited the Fermi National
Accelerator Laboratory and he is grateful to the Fermilab Theory Group
for their hospitality. {\em This work was supported in part by the
  Department of Energy under grant DE-FG02-91ER40676 and the National
  Science Foundation under grant PHY-9218167.}


\bibliography{align}
\bibliographystyle{h-elsevier}

\end{document}